# Feasibility Study of Hybrid Inverse Planning with Transmission Beams and Single-energy Spread-out Bragg Peaks for Proton Flash Radiotherapy

**Chaoqiong Ma, Xiaofeng Yang[*], Chih-Wei Chang, Ruirui Liu, Duncan Bohannon, Liyong Lin, Tian Liu, Sibo Tian and Jun Zhou[*]**

Department of Radiation Oncology, Emory University, Atlanta, GA, 30322, USA

[*]Authors to whom correspondence should be addressed.

Email: xiaofeng.yang@emory.edu and jun.zhou@emory.edu

**Abstract**

**Purpose**: Ultra-high dose rate (FLASH) proton planning with only transmission beams (TBs) has limitations in normal tissue sparing. The single-energy spread-out Bragg peaks (SESOBPs) of FLASH dose rate have been demonstrated feasible for proton FLASH planning. We investigated the feasibility of combining TBs and SESOBPs for proton FLASH treatment.

**Methods**: A hybrid inverse optimization method was developed to combine the TBs and SESOBPs (TB-SESOBP) for FLASH planning. The SESOBPs were generated field-by-field from spreading out the BPs by pre-designed general bar ridge filters (RFs) and placed at the central target by range shifters (RSs) to obtain a uniform dose within the target. The SESOBPs and TBs were fully sampled field-by-field allowing automatic spot selection and weighting in the optimization process. A spot reduction strategy was conducted in the optimization process to push up the minimum MU/spot assuring the plan deliverbility at beam current of 165 nA. The TB-SESOBP plans were validated in comparison with the TB only (TB-only) plans and the plans with the combination of TBs and BPs (TB-BP plans) regarding 3D dose and dose rate (dose-averaged dose rate) distributions for five lung cases. The FLASH dose rate coverage ($V_{40Gy/s}$) was evaluated in the structure volume received >10% of the prescription dose.

**Results**: Comparing to the TB-only plans, the mean spinal cord $D_{1.2cc}$ drastically reduced 41% ($P < 0.05$), the mean lung $V_{7Gy}$ and $V_{7.4Gy}$ moderately reduced by up to 17% ($P < 0.05$) and the target dose homogeneity slightly increased in the TB-SESOBP plans. Comparable dose homogeneity was achieved in both TB-SESOBP and TB-BP plans. Besides, prominent improvements were achieved in lung sparing for the cases of relatively large targets by the TB-SESOBP plans comparing to the TB-BP plans. The targets were fully covered with the FLASH dose rate in all the three plans. For the OARs, $V_{40Gy/s}$ = 100% was achieved by the TB-only plans while $V_{40Gy/s}$ > 85% was obtained by the other two plans.

**Conclusions**: We have demonstrated that the hybrid TB-SESOBP planning was feasible to achieve FLASH dose rate for proton therapy. With pre-designed general bar RFs, the hybrid TB-SESOBP planning can be potentially implemented for proton adaptive FLASH radiotherapy. As an alternative FLASH planning approach to TB-only planning, the hybrid TB-SESOBP planning has great potential in improving OAR sparing while maintaining high target dose homogeneity.

1. Introduction

Ultrahigh dose rate irradiation, also known as FLASH radiotherapy (RT) and generally considered as mean dose rate ≥ 40 Gy/s, has been demonstrated to reduce normal tissue toxicities while retaining anti-tumor efficacy comparing to the conventional dose-rate irradiation[1–5]. Normal tissue sparing by the FLASH effect has been observed in several *in vivo* small animal studies using electron, photon and proton beams[1–8]. The radiobiological mechanisms contributing to the FLASH effect are not yet been fully clarified, however, the oxygen depletion and reactive oxygen species production are the most current valid hypothesis.[9,10] The FLASH treatment of a first patient with T-cell cutaneous lymphoma using electron beam has further confirmed the technical feasibility and clinical safety of the FLASH-RT, as well as its promising outcome both on normal skin and the tumor[11]. Additionally, a clinical trial of applying FLASH-RT for the treatment symptomatic bone metastases is underway[12]. The encouraging findings in these preclinical studies provide a rationale for translating the FLASH-RT into future clinical RT.

The FLASH delivery using proton beams has attracted intensive attention recently owning to the much larger penetration depth of the protons relative to the electrons, which enables the irradiation of the deep-seated tumors[9]. Besides, the low entrance dose and the deposition of majority energy in a narrow range, known as Bragg peak (BP), of the proton beams prompt more accurate targeting of the tumor and improved normal tissue sparing in contrast to the photon beams. Pencil beam scanning (PBS) has been widely implemented for intensity-modulated proton therapy (IMPT) in which the spread-out BPs (SOBPs) are generated using multiple energy layers to provide highly conformal dose to the target[13]. However, the conventionally utilized clinical cyclotron system for PBS requires energy degraders, energy selection and energy switch system to obtain desired proton energies, which drastically reduces the efficiency and the proton beam current and thereby prevents the conventional IMPT planning from achieving FLASH dose rate[14]. To overcome this difficulty, attempts have been conducted in adopting proton beams of a constant high energy for FLASH planning. Generally, the previous studies have used shoot-through method to cover the target by the entrance dose region of the high-energy proton transmission beams (TBs) and deliver the BPs outside the patient, which have been indicated feasible to obtain FLASH effect[15–18]. One major drawback of using TBs alone (TB-only) for FLASH planning is the over irradiation exposure of normal tissues located beyond the distal edge of target. Thus, a method with the joint use of the TBs and the BPs of multi-energy proton beams was developed in the further study by placing BPs inside the target for target dose conformality and normal tissue sparing while using TBs to cover the target boundary and maintain FLASH dose rate coverage for the adjacent organs-at-risk (OARs)[19].

Another superiority of the conventional proton RT to the photon and electron RT is the enhanced relative biological effect (RBE) owing to the utilization of the BPs with high linear energy transfer (LET) effect for the target dose coverage[20]. However, such advantage of the proton beams could not be exploited by the TB-only planning. Moreover, the LET effect could potentially be boosted by the BPs with FLASH dose rate to further improve tumor control as well as reduce toxicity[21]. The feasibility of using single-energy BPs with FLASH dose rate for planning has been explored in recent studies[22,23]. In these studies, the uniform range shifters (RSs) and patient-specific universal range compensators were employed to align the BPs of the high-energy proton beams to the distal edge of the target from multiple beam directions, which demonstrated improved OAR sparing with sufficient target dose coverage and comparable FLASH dose rate coverage comparing to the TB-only planning. Alternative approaches have been proposed for FLASH planning using spread-out single-energy proton beams, in which patient-specific range compensators were used to pull back the BPs to the target exit edge and pin-shaped ridge filters (RFs) were customized to spread out the BPs to the proximal edge of the target[24,25]. Nevertheless, either using the customized range compensators alone or combining the patient-specific pin-shaped RFs and range compensators for FLASH planning has limitations in adapting to patient anatomical changes such as tumor regression, which often occur during the treatment courses, since a time-consuming process for re-making the range compensators and/or pin-shaped RFs may be required.

Instead of customizing the SOBPs of the proton pencil beams regarding the target edges as the patient-specific pin-shaped RFs, general bar RFs consist of multiple identical RF bars that take the monoenergetic proton beam with varying material thicknesses to obtain uniform dose distribution over a specified wide range of depth, referred to as single-energy SOBPs (SESOBPs)[8,26]. The fixed plateau width of the SESOBPs is determined by the height of the RF bars. Although the pre-designed general bar RFs of varying heights could be potentially implemented to generate the SESOBPs of various plateau widths regarding the target sizes, using only SESOBPs of fixed plateau width could hardly achieve conformal dose in the irregular shaped targets. Being inspired by the aforementioned joint TB and BP planning, in this study, we proposed a hybrid inverse planning method to combine SESOBPs generated by pre-designed general bar RFs and TBs for adaptive proton FLASH planning, aiming to place SESOBPs inside the targets to fully exploit its high FLASH and LET effect while using TBs to account for the dose conformality at the target boundaries and maintain FLASH dose rate coverage for the surrounding normal tissues.

## 2. Methods and materials

### 2.1 Overview

This study was conducted using monoenergetic proton beams of the highest energy 250 MeV generated by a Varian ProBeam system in the "FLASH mode". The dose calculation for treatment planning was based on Monte Carlo (MC) simulations using a fast open-source MC tool, MCsquare[27], which has been commissioned using a single Gaussian beam source model based on the measured integrated depth dose (IDD) and in-air lateral profiles of 250 MeV beams delivered by the ProBeam system in our facility[28]. Excellent agreements of -0.01 mm $R_{80}$ (80% dose falloff) and -0.31 mm BP width (full width at half maximum, FWHM) difference between MCsquare calculated and measured IDD were observed.

A set of general bar RFs were pre-designed aiming to form SESOBPs from 250 MeV beams with plateau widths ranging from 30 to 70 mm. Each RF was comprised of multiple identical discrete-step-shaped bar ridges with the height and width of each step being optimized to minimize dose fluctuations within the plateau region (see Figure 1(b)).

A hybrid inverse optimization method was developed based on optimization of the beam intensities with a combination of TBs and SESOBPs (hybrid TB-SESOBP) for FLASH planning. To generate SESOBPs for each field that could cover the inside of the target, the BPs were pulled to a depth nearby the distal edge of the target using a uniform range shifter and spread out within the target by a selected pre-designed general bar RF. The SESOBPs and TBs were fully sampled to cover the target field-by-field (Figure 1(a)) which could allow automatic spot selection and weighting in the hybrid inverse optimization process. To enable FLASH dose rate for beam delivery, a spot reduction strategy was conducted in the optimization process by iteratively reducing the low weighted spots and optimizing the weights of the residual spots, thereby pushing up the monitor unit (MU)/spot above the minimum deliverable MU in the "FLASH mode".

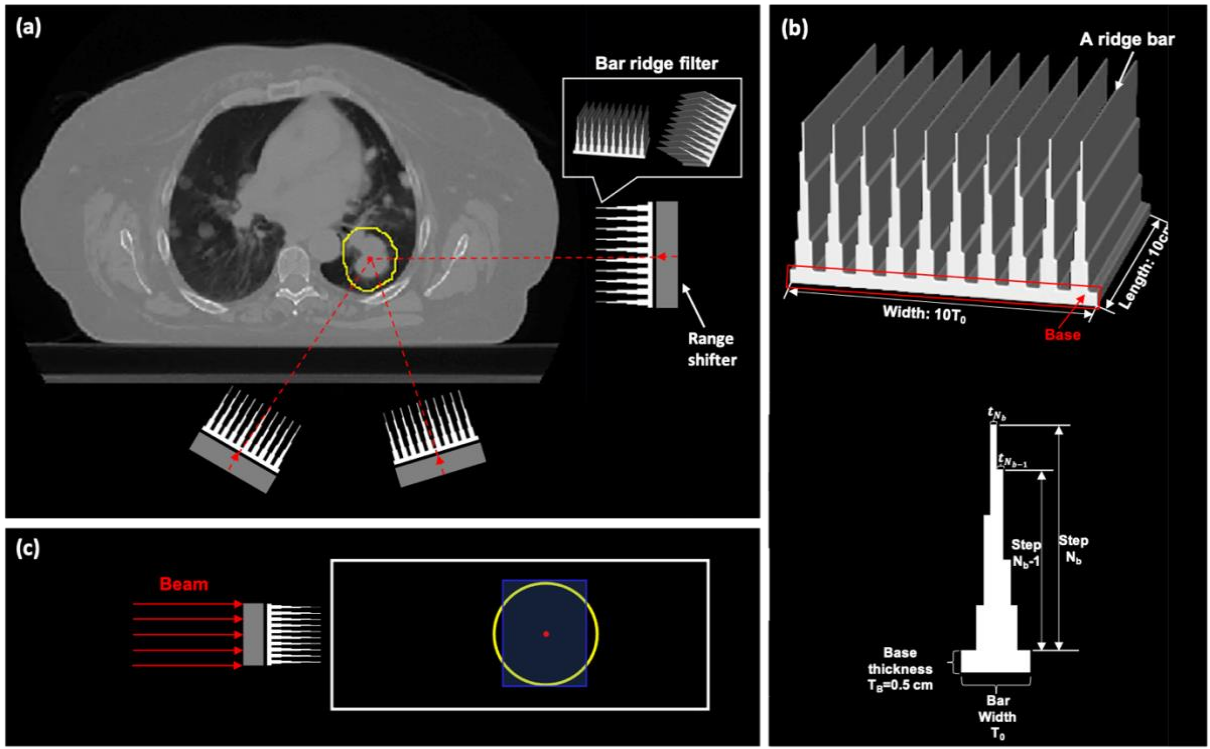

**Figure 1**. (a) A three-field arrangement for a lung treatment planning using RSs and general bar RFs, with two 3D views of one bar RF. The target is contoured in yellow. (b) 3D rendering of a general bar RF with 10 equally spaced identical ridge bars and a base (top), and the cross-sectional shape of a ridge bar (bottom). (c) An example of the expected plateau region (rectangle outlined in blue) of the SESOBPs generated by a RS and RF in a water phantom with a spherical target (contoured in yellow).

## 2.2 Design of general bar RFs

The mechanism of spreading out the BPs of an monoenergetic proton beam using a step-shaped bar RF is to modulate the ranges, referred to as water equivalent depths (WED) of the BPs, and the fluences of the incident protons by the steps of varying thicknesses and widths, thereby a SESOBP of larger and fixed plateau width can be formed by the weighted combination of the Bragg curves of varying WEDs of the BPs[26]. A general bar RF consisted of multiple equally spaced identical step-shaped bar ridges mounting to a base is shown in Figure 1 (b). For an RF with ridges of $N_b$ steps, Bragg curves of $N_b$ different WEDs of the BPs can be resulted from an impinging monoenergetic proton beam. Let $R_j$ denote the WED of the $j^{th}$ BP, $j = 1, ..., N_b$ and $R_1 > ... > R_{N_b}$. Thus, the width of the SOBP plateau region equals to the difference between $R_1$ and $R_{N_b}$, which corresponded to Bragg curves resulted from the modulation by the thinnest and the thickest steps of the bar ridges, respectively. Let $z_i$ denote the position of the $i^{th}$ point of $N_p$ uniformly spaced points in the plateau region of a SESOBP and $B_j(z_i)$ denote the dose at $z_i$ from the $j^{th}$ Bragg curves. The dose at $z_i$, $D(z_i)$, for the SESOBP, which is the weighted superposition of the Bragg curves at $z_i$, can be expressed as follows,

$$D(z_i) = \sum_j^{N_b} \omega_j B_j(z_i) \qquad (1)$$

where $\omega_j$ is the weight of the $j^{th}$ Bragg curve, which is determined by the proportion of the beam modulated by the corresponding step and thereby proportional to the step width, $t_j$. Therefore, to obtain homogeneous doses in the SESOBP plateau region, the least squares regression problem in formula (2) can be solved by finding the optimal weights to minimize the sum of squared errors between the point doses of the SESOBP plateau region and the desired does, $D_0$. The optimal weights, $\omega^*$, can then be translated to the step widths by equation (3), where $T_0$ is the ridge bar width and equals to the summation of the step widths.

$$\min_\omega \sum_i^{N_p} ||D_0 - \sum_j^{N_b} \omega_j B_j(z_i)||^2 \qquad (2)$$

$$s.t., \omega_j \geq 0$$

$$t_j = \frac{\omega_j^*}{\sum_j^{N_b} \omega_j^*} \cdot T_0 \qquad (3)$$

In this study, a set of nine general bar RFs of water equivalent heights ranging from 3 to 7 cm with a spacing of 0.5 cm were pre-designed aiming to form SESOBPs from 250 MeV beams with plateau widths of equal heights of the corresponding RFs. The material density of 1.0 g/cm³ was used for the RFs in the simulations. Therefore, the physical height of a ridge bar equaled to its water equivalent height, which is shown in Figure 1(b) as the height of the step $N_b$. The width of each ridge bar $T_0$ was set to 10 mm and the bar length was set to 10 cm. Each RF consist of ten equally spaced ridge bars mounted to a base with thickness of 5 mm and the same length of the ridge bars. Thus, the width of the base was the total width of ten ridge bars. To ensure high computational efficiency including high calculation speed and reasonable memory consumption, as well as the high precision in RF design, the RFs were simulated with a voxel resolution of $1 \times 1 \times 1$ mm³. A set of 71 Bragg curves of equally spaced BP positions for the 250 MeV were acquired by calculating the IDDs of a pencil beam in a water phantom after modulated by homogeneous RSs with water equivalent thicknesses (WETs) ranging from 0 to 70 mm. The optimized combination of the Bragg curves that could form the SESOBPs of minimized dose fluctuations were obtained by solving the problem in formula (2) and then the thicknesses as well as the translated widths of the corresponding RSs were used to design each bar RF.

## 2.3 Generation of SESOBPs

As the core of the proposed methodology for FLASH planning, the utilization of the SESOBPs aims to place the SESOBP plateau regions of high LET effect inside of the targets so that improved therapeutic ratio could be achieved. The SESOBPs were generated by the combination of a RS and a selected general bar RF from the pre-designed RF set, in which the BPs of the scanning pencil beams were pulled to a same WED near the distal edge of the target by the RS and spread out by the RF to a depth close to the proximal edge of the target along the beam direction. The pencil beam spots were fully sampled over the target at each field. Thus, for a spherical target in a homogeneous phantom, uniform dose could be formed by the plateau regions of the SESOBPs in a cuboid region, as is shown in Figure 1(c), which could hardly conform to irregular shaped targets. To generate SESOBPs that could cover the inside of the target meanwhile mitigating the plateau region of the SESOBPs located beyond the target, we chose the 90th percentile WEDs of the distal edges and the 10th percentile WEDs of the proximal edges of the target along the pencil beam directions as the ending, $R_e$, and beginning, $R_b$, of the plateau region, respectively. To pull the 90% dose falloff, $R_{90}$, of the 250 MeV Bragg curves back to $R_e$, the WET of the RS, $T_{RS}$, can be calculated by equation (4). $T_B$ is the WET of the RF base, which served to pull back the BP positions by $T_B$ (0.5 cm was used). The bar RF of height less than and closest to $R_e - R_b$ were selected to generate SESOBPs within the expected region. Of particular note that the RF was attached to the bottom of the RS and positioned as close as possible to the patient surface to reduce the in-air scattering and preserve high beam concentration.

$$T_{RS} = R_{90} - R_e - T_B \qquad (4)$$

**2.4 Hybrid TB-SESOBP optimization for FLASH planning**

*2.4.1 Hybrid TB-SESOBP optimization problem*

Although the generated SESOBPs could potentially improve tumor control with high LET effect and mitigate dose delivered to the normal structures located beyond the distal edge of the target comparing to the TBs, high dose uniformity in the target could hardly be achieved by planning with the SESOBPs only due to its cuboid-shaped plateau region centered in the target. Therefore, in the proposed hybrid planning method for FLASH planning, we employed the TBs as the compensation for SESOBPs to primarily cover the target boundary. Similarly, the TBs were fully sampled to cover the target at each field, which could allow more degrees of freedom in the optimization for automatic beam combination selection to further improve plan quality. Let $M_B$ and $M_S$ denote the numbers of spots, $w^B$ and $w^S$ denote spot weights, $d^B$ and $d^S$ denote the dose influence matrixes for the TBs and SESOBPs, respectively. Thus, the hybrid TB-SESOBP planning can be formulated in (5) as a constrained optimization problem, in which the spot weights of the TBs and SESOBPs can be optimized to obtain homogeneous dose in the target meanwhile minimize normal structure toxicity. Here $\rho_o$ and $\rho_T$ are the penalties, $N_O$ and $N_T$ are the numbers of voxels, $\bar{D}^o$ and $\bar{D}^T$ are the reference doses, for an OAR and a target, respectively. The dose distribution, $D$, obtained from the hybrid planning is the summation of the dose components from TBs and SESOBPs.

$$\min_{(w^B, w^S)} f = \sum_{o \in OAR} \frac{\rho_o}{N_O} \sum_{i=1}^{N_O} \{\max(0, D_i - \bar{D}_i^o)\}^2 + \sum_{T \in Targets} \frac{\rho_T}{N_T} \sum_{i=1}^{N_T} (D_i - \bar{D}_i^T)^2$$

$$s.t. \quad D_i = \sum_{j=1}^{M_B} w_j^B d_{ij}^B + \sum_{j=1}^{M_S} w_j^S d_{ij}^S \qquad (5)$$

$$w^B \geq 0, w^S \geq 0$$

*2.4.2 Iterative spot reduction (ISR) process*

Table 1. ISR process

| ISR Process |
| --- |
| **Begin** |
| Set minimum spot weight $w_{min}$ |
| Set maximum number of spots, $N_M$, can be removed at each iteration |
| Set spot weights $w = \{w_1^B, \dots, w_{M_B}^B, w_1^S, \dots, w_{M_S}^S\}$ |
| Initialize the number of spots, $N_R$, to be removed as 0 |
| **Repeat** |
|     Sort spots by weight in ascending order |
|     Remove the first $N_R$ spots from planning and the corresponding weights, $w_R$, from $w$, $w = w/w_R$ |
|     Get optimal spot weights $w^* = \arg\min_w f$ and set $w = w^*$ |
|     Get the number of spots, $N_s$, of weights $w_{min}$ |
|     $N_R = \min(N_M, N_s)$ |
| **Until** $N_R = 0$ |
| **Output** $w$ |
| **End** |

According to the previous studies[29, 30], the proton dose rate of an energy layer is proportional to the minimum MU/spot (minimum spot weight) at a fixed minimum spot duration time in a Varian ProBeam system. Thus, the minimum spot weight needs to be above a certain threshold to ensure that the plan is deliverable at FLASH dose rate. For instance, the minimum deliverable spot weight is approximately 515 MU, corresponding to around $5.15 \times 10^8$ protons, given a beam current of 165 nA and a minimum spot time of 0.5 ms. In this study, a spot reduction strategy was conducted in the optimization process by iteratively removing the low weighted spots and optimizing the weights of the residual spots, thereby pushing up the MU/spot above the minimum deliverable MU, denoting as $w_{min}$, at FLASH dose rate. The pseudocode of the spot reduction process is listed in Table 1. Specifically, the set of spot weights, $w$, was initialized with the weights of all the $M_B + M_s$ spots of TBs and SESOBPs and the number of spots, $N_R$, to be removed was initialized as 0. A maximum number of spots to be removed each time was limited as $N_M$ prior to the spot reduction process. In each iteration, the $N_R$ least weighted spots were firstly removed from the planning and the corresponding weights, $w_R$, were excluded from $w$. Then, $w$ were updated by the optimal weights, $w^*$, acquired from solving the optimization problem formulated in (5). At last, the number of spots, $N_s$, of weights $\geq w_{min}$ was obtained and the minimum value between $N_R$ and $N_M$ was assigned to $N_R$. The process terminated until $N_R = 0$, in which case the weights of all the residual spots were above $w_{min}$. Note that $N_M$ was set as 10% of the total number of the initial spots considering efficiency and effectiveness of the optimization process.

### 2.4.3 *The hybrid optimization framework*

An optimization framework based on an open-source toolkit 'matRad'[31] was established to generate the hybrid TB-SESOBP plans. In this framework, the CT and RTStruct files in DICOM format as well as beam parameters including gantry angles and beam energies (250 MeV) were taken as the input to obtain the information of the RSs and bar RFs for generating the SESOBPs, the incorporation of which was fed into MCsquare to generate the dose influence matrixes. Given the dose influence matrixes, prescriptions of the targets and the dose constraints of the OARs, the interior point optimizer package (IPOPT)[32] provided by matRad was then employed in conducting the spot reduction process to repeatedly solve the optimization problem defined in (5), until the deliverable spot weights at FLASH dose rate were acquired.

## 2.5 Patient study

The hybrid TB-SESOBP planning method was validated on five lung cancer patients of clinical target volume (CTV) volume 85±43 cc who underwent proton stereotactic body radiation therapy (SBRT) at our institution. For this study, the prescription dose was set to 34 Gy delivered in a single fraction[33,34] and the OAR dose constraints were adopted from the corresponding RTOG 0915 protocol. Similar to the conventional IMPT planning in our clinic, three beam angles were empirically chosen for each case to ensure target dose conformality and normal structure avoidance.

In this study, the dose averaged dose rate (DADR)[16] metric was employed to quantify the voxel dose rate of each plan. According to the previous study based on the Varian ProBeam system[29], the beam current was set to 165 nA and a relatively small achievable minimum spot time of 0.5 ms was chosen, which could allow a lower minimum spot weight in planning and thereby improve plan quality. Therefore, the minimum spot weight $w_{min}$ in the FLASH planning was 515 MU corresponding to $5.15 \times 10^8$ protons.

For comparison of the FLASH dose rate, a TB-only FLASH plan was generated for each case through the proposed ISR process. Additionally, the ISR process was conducted to obtain a FLASH plan with the combination use of TBs and BPs (hybrid TB-BP) for each case serving as the reference to assess the effects of the bar RFs in the hybrid TB-SESOBP planning. Note that the BPs in the hybrid TB-BP planning were generated by only removing the bar RFs in the generation of the SESOBPs. The beam angles, prescription dose and OAR dose constraints used in the TB-only and the hybrid TB-BP plans were kept consistent with those in the corresponding hybrid TB-SESOBP plans. The 3D dose distributions and the 3D dose rate distributions of the three plans for each case were evaluated and the hybrid TB-SESOBP plan was compared to the other two plans in terms of the dose volume histograms (DVHs) and the dose rate volume histograms (DRVHs). Note that DRVHs of each plan were evaluated in the volume receiving clinically significant dose of ≥10% of the prescription.

## 3. Results

### 3.1 Pre-designed bar RFs

**Table 2**. Step thicknesses and widths (in parenthesis) of the pre-designed bar RFs with heights ranging from 30 to 70 mm.

| RF height (mm) | 30 | 35 | 40 | 45 | 50 | 55 | 60 | 65 | 70 |
|---|---|---|---|---|---|---|---|---|---|

| Step thickness (width) (mm) | 0 (6), 10 (1), 15 (1), 20 (1), 30 (1) | 0 (5), 7 (1), 14 (1), 21 (1), 28 (1), 35 (1) | 0 (5), 10 (2), 20 (1), 30 (1), 40 (1) | 0 (5), 10 (1), 15 (1), 25 (1), 35 (1), 45 (1) | 0 (5), 10 (1), 15 (1), 25 (1), 40 (1), 50 (1) | 0 (5), 11 (2), 22 (1), 33 (1), 44 (1), 55 (1) | 0 (5), 12 (2), 24 (1), 36 (1), 48 (1), 60 (1) | 0 (5), 13 (2), 26 (1), 39 (1), 52 (1), 65 (1) | 0 (5), 14 (2), 28 (1), 42 (1), 56 (1), 70 (1) |
|---|---|---|---|---|---|---|---|---|---|

The parameters of each pre-designed RF including the heights and widths of the steps are listed in Table 2. The optimized ridge bars were composed of five to six steps of thicknesses ranging from 0 to the height of the corresponding ridge bar. For each ridge bar, the step of thickness 0, which located at the bottom of the ridge bar corresponding to the original BP, had the largest width of 5-6 mm. The widths of the rest steps were 1-2 mm. The resulted widest step of thickness 0 could be explained by the reduced overlapping of the modulated Bragg curves with the increased depth, especially at the tail of the SESOBP plateau region, which was mainly contributed by the original BP. It is important to note that the step widths listed in Table 2 were firstly derived from the equation (3) and then rounded to the nearest mm due to the limitation of the voxel resolution in simulations. Thus, for the pre-designed bar RFs of heights ranging from 55 to 70 mm, of which the summations of the step widths were 11 mm (> 10 mm, the pre-set bar width $T_0$), the bar widths were set to 11 mm and the RF widths were 11 cm accordingly to maintain the weightings of the steps for modulations.

The rounded step widths could affect the flatness of the SESOBPs. Figure 2 (a) illustrates the SESOBPs with plateau widths of 30, 50 and 70 mm. To quantify the flatness of each SESOBP, the SESOBP was normalized to its maximum depth dose and the mean as well as the standard deviation (std) of the percent dose in the plateau region were calculated. As plotted in Figure 2 (b), the SESOBPs with plateau widths of 30 and 40 mm had the highest flatness: the mean percent doses and the stds were $96 \pm 2\%$. Followed by the SESOBPs with plateau widths of 35, 45 and 50 mm, in which the mean percent doses were around 93% and the stds were around 3.5-4%. The plateau flatness reduced with increase of the plateau width: as the plateau width increased from 55 to 70 mm the mean percent dose decreased from $92.2 \pm 3.4\%$ to $89.1 \pm 4.1\%$. Therefore, higher dose fluctuations were introduced by the resolution limitation on the SESOBP plateau regions of widths larger than 50 mm.

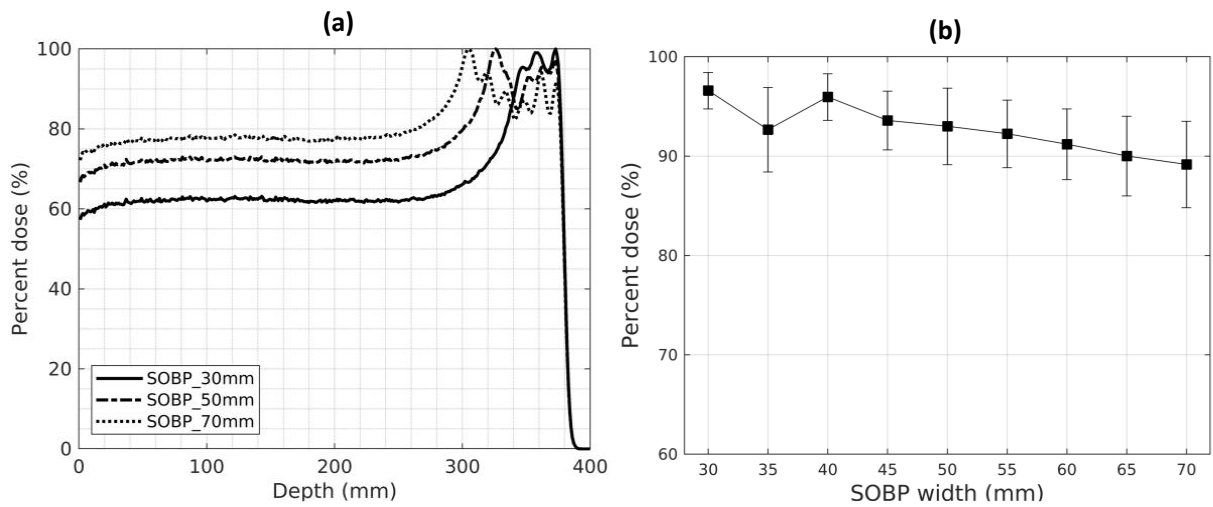

**Figure 2**. (a) SESOBPs of plateau widths 30, 50 and 70 mm; (b) means (square markers) and stds (error bars) of the percent dose in the SESOBP plateau regions of widths ranging from 30 to 70 mm.

## 3.2 Dosimetric evaluation

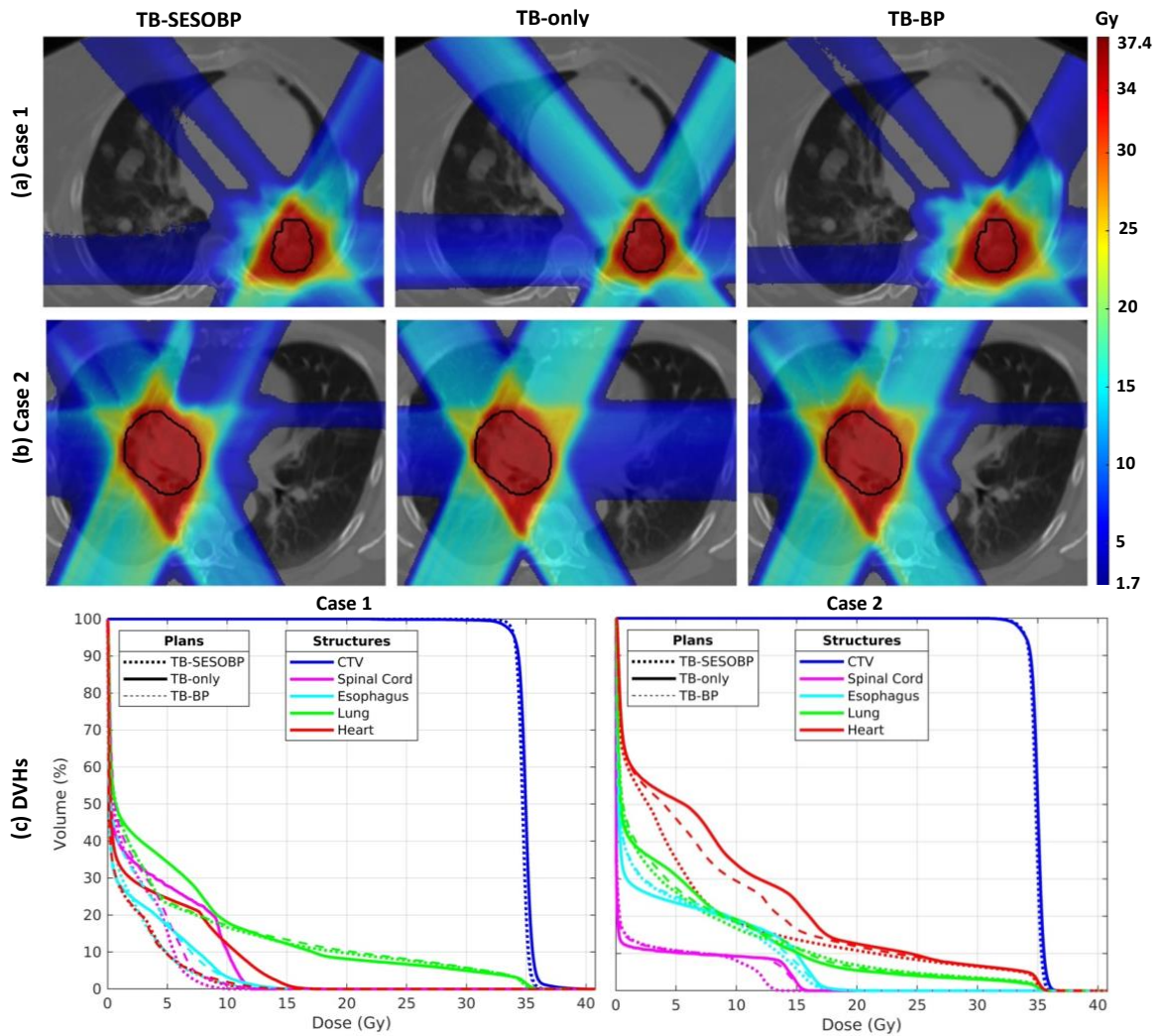

**Figure 3**. 2D dose distributions of the three plans for (a) Case 1 and (b) Case 2. The CTVs are contoured in black. (c) DVH comparisons between the three plans for the two cases. The DVHs of the TB-SESOBP, TB-only and TB-BP plans are marked in dotted, solid and dashed lines, respectively. The dose plot window is [1.7 Gy, 37.4 Gy].

Table 3. Dosimetry indices of the TB-SESOBP, TB-only and TB-BP plans for the five cases

| | | Lung | | Spinal cord | | | Esophagus | | Heart | | CTV | |
|---|---|---|---|---|---|---|---|---|---|---|---|---|
| | | $V_{7Gy}$ (cc) | $V_{7.4Gy}$ (cc) | $D_{0.35cc}$ (Gy) | $D_{1.2cc}$ (Gy) | $D_{max}$ (Gy) | $D_{5cc}$ (Gy) | $D_{max}$ (Gy) | $D_{15cc}$ (Gy) | $D_{max}$ (Gy) | $D_{2\%}$ (%) | Volume (cc) |
| Case 1 | TB-SESOBP | 225.4 | 220.1 | 6.9 | 5.4 | 10.5 | 2.3 | 13.5 | 8.1 | 13.9 | 105.3 | 45.1 |
| | TB-only | 313.8 | 299.0 | 11.1 | 9.9 | 12.2 | 2.8 | 16.0 | 13.6 | 18.9 | 108.1 | 45.1 |
| | TB-BP | 229.8 | 223.2 | 8.2 | 6.1 | 13.2 | 4.0 | 14.6 | 8.6 | 17.2 | 104.7 | 45.1 |
| Case 2 | TB-SESOBP | 457.1 | 442.3 | 12.9 | 12.4 | 13.6 | 10.8 | 21.6 | 34.8 | 36.4 | 105.0 | 92.7 |
| | TB-only | 527.6 | 502.5 | 15.5 | 15.1 | 16.2 | 11.9 | 22.1 | 35.1 | 36.5 | 106.1 | 92.7 |
| | TB-BP | 488.9 | 472.7 | 15.4 | 14.9 | 16.2 | 11.4 | 21.3 | 35.0 | 36.9 | 105.5 | 92.7 |
| Case 3 | TB-SESOBP | 469.0 | 451.5 | 11.2 | 9.2 | 12.5 | 13.8 | 38.7 | 3.4 | 13.5 | 112.3 | 129.1 |
| | TB-only | 543.9 | 534.7 | 12.1 | 8.5 | 14.0 | 14.1 | 38.8 | 0.7 | 6.7 | 114.1 | 129.1 |
| | TB-BP | 498.9 | 489.1 | 8.3 | 7.3 | 9.6 | 14.1 | 37.8 | 3.6 | 14.7 | 113.6 | 129.1 |
| Case 4 | TB-SESOBP | 243.5 | 239.4 | 3.9 | 2.9 | 5.6 | 1.8 | 4.7 | 10.9 | 30.6 | 105.2 | 72.7 |
| | TB-only | 302.4 | 298.6 | 12.7 | 11.8 | 13.9 | 10.4 | 12.7 | 12.0 | 31.0 | 118.7 | 72.7 |
| | TB-BP | 255.3 | 250.7 | 1.9 | 1.3 | 3.5 | 1.7 | 4.6 | 14.2 | 30.7 | 105.2 | 72.7 |
| Case 5 | TB-SESOBP | 108.9 | 107.0 | 6.8 | 5.2 | 12.1 | 9.5 | 25.9 | 0.1 | 0.3 | 106.7 | 42.0 |
| | TB-only | 116.5 | 112.1 | 14.8 | 13.6 | 16.4 | 12.7 | 27.3 | 0.1 | 0.3 | 117.0 | 42.0 |
| | TB-BP | 103.4 | 101.0 | 12.4 | 9.0 | 16.0 | 9.5 | 26.0 | 0.1 | 0.3 | 106.1 | 42.0 |
| | TB-SESOBP | 300.8±156.9 | 292.1±150.2 | 8.3±3.6 | 7.0±3.8 | 10.9±3.1 | 7.6±5.3 | 20.9±12.8 | 11.5±13.7 | 18.9±14.5 | 106.9±3.1 | |
| | TB-only | 360.8±177.9 | 349.4±172.6 | 13.4±1.9 | 11.8±2.7 | 14.5±1.8 | 10.4±4.4 | 23.4±10.3 | 12.3±14.2 | 18.7±15.4 | 112.8±5.5 | |
| | TB-BP | 313.1±170.2 | 305.3±165.6 | 9.1±5.1 | 7.6±4.9 | 11.6±5.4 | 8.1±5.2 | 21.0±12.6 | 12.3±13.8 | 20.0±14.3 | 107.0±3.7 | |
| $P$ value | TB-SESOBP vs. TB-only | **0.01** | **0.01** | 0.06 | **0.04** | 0.05 | 0.15 | 0.16 | 0.56 | 0.90 | 0.08 | |
| | TB-SESOBP vs. TB-BP | 0.13 | 0.13 | 0.67 | 0.65 | 0.65 | 0.20 | 0.78 | 0.26 | 0.15 | 0.76 | |

The 2D dose distributions of the three plans for Case 1 and 2 are illustrated in Figure 3 (a) and (b), respectively, and the corresponding DVHs are illustrated in Figure 3 (c). In general, reduced exit dose in the normal tissues near the distal edge of the targets along each beam direction was observed in the TB-SESOBP and TB-BP plans comparing to the TB-only plans. Moreover, the TB-SESOBP plan of Case 2, which had a larger CTV comparing to Case 1 (see Table 3), yield less exit dose than the TB-BP plan, leading to reduced dose to the heart and esophagus according to the DVH comparisons. The exit dose of the two hybrid beam plans for Case 1 was comparable. To further quantify the plan quality, the dosimetry indices of the concerned OARs according to the RTOG 0915 protocol were extracted for each plan and are listed in Table 3. To assess the target dose homogeneity, the dose delivered to 2% of the target volume ($D_{2\%}$) was calculated for each plan and listed in Table 3 as well. In addition, these dosimetry indices were analyzed statistically between the hybrid TB-SESOBP and the other two plans using paired $t$-test with a significance level of 5% ($p < 0.05$). The TB-SESOBP plans achieved statistically significant improvements in spinal cord and lung sparing compared to the TB-only plans. Specifically, the mean spinal cord $D_{1.2cc}$ drastically decreased from 11.8 Gy for the TB-only plans to 7.0 Gy for the TB-SESOBP plans, corresponding to a reduction of 41%. The mean $V_{7Gy}$ and $V_{7.4Gy}$ of the lung drastically dropped by 60 and 57 cc, respectively, from the TB-only plans to the TB-SESOBP plans, corresponding to a reduction up to 17%. In terms of target dose homogeneity, the mean $D_{2\%}$ of the two types of the hybrid beam plans were nearly identical, which were reflected in the DVH comparisons in Figure 3 (c). Though no statistically significant differences in $D_{2\%}$ between the hybrid beam plans and the TB-only plans, lower $D_{2\%}$ was observed in the hybrid beam plans than the TB-only plan for every case and this difference was up to 13.5%. The higher target dose homogeneity achieved by the hybrid beam plans could be resulted from the increased optimization degrees of freedom in planning. Besides, the TB-SESOBP plans had no statistical superiority in OAR sparing over the TB-BP plans.

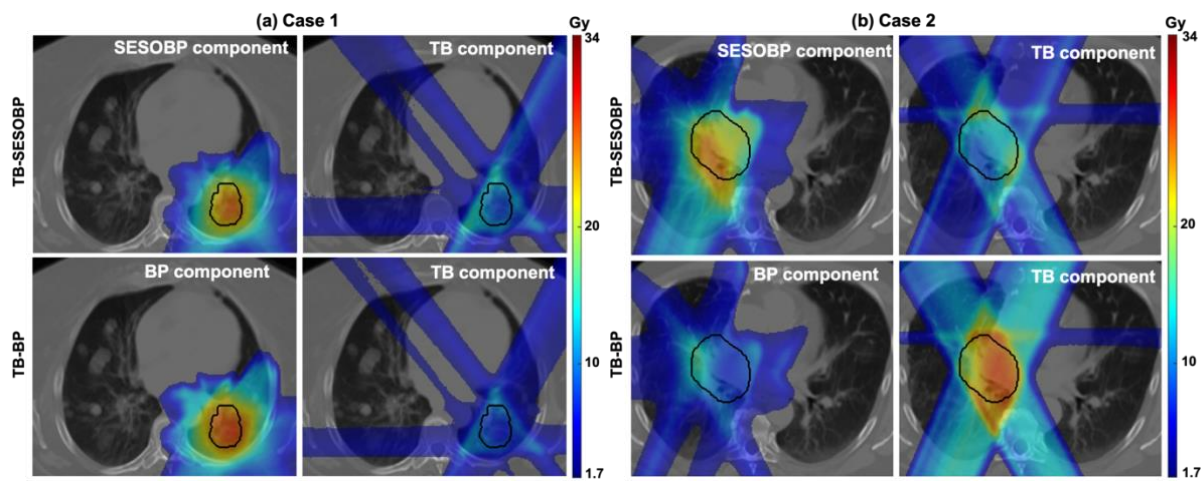

**Figure 4**. SESOBP component (top left) and TB component (top right) of the TB-SESOBP plan, and BP component (bottom left) and TB component (bottom right) of the TB-BP plan for (a) Case 1 and (b) Case 2. The CTVs are contoured in black. The dose plot window is [1.7 Gy, 34 Gy].

To gain some insights into the impacts brought by the RFs in planning, we decomposed the dose distribution of each TB-SESOBP plan into TB and SESOBP dose components and compared them with the TB and BP dose components of the corresponding TB-BP plan. Figure 4 illustrates the decomposed dose components of the two hybrid beam plans for Case 1 and 2. As expected, the dose of the SESOBP component primarily concentrated in the central target for protection of the OARs beyond the target meanwhile the TB dose component was mainly weighted at the edge of the target to assure high dose coverage to the target in the TB-SESOBP planning for both cases. Additionally, higher dose was observed in the target boundary for the TB component and in the central target for the BP component of the TB-BP plan of Case 1. However, such dose distribution patterns were not shown in the TB-BP plan of Case 2 with larger CTV. Besides, the proportion of the TB component of the TB-BP plan was larger than that of the TB-SESOBP plan for Case 2 whereas similar weighting of the TB component was observed in the two hybrid beam plans for Case 1. The relatively high weighting of the TB component in the TB-BP plan for Case 2 could be resulted from the limited width of the BPs since the TBs were needed not only to preserve dose coverage at the target boundary but also to compensate the insufficient dose in the central target. This could explain the superiority of the TB-SESOBP plans over the TB-BP plans for the cases of relatively large targets. For instance, prominent improvements were achieved in lung sparing from employing RFs in planning for Case 2 and 3 whereas the discrepancy of the plan quality between the two hybrid beam plans for Case 1 and 5 was negligible. It is worth mentioning that the benefits for the heart, esophagus and spinal cord gained from involving RFs in the hybrid beam planning were correlated to the spatial relationship between the target and these OARs as well as beam arrangements. For the OARs located near the distal edge of the target along the beam directions of relatively high weighting in the whole plan, better protection could be obtained from involving SESOBPs.

### 3.3 Dose rate evaluation

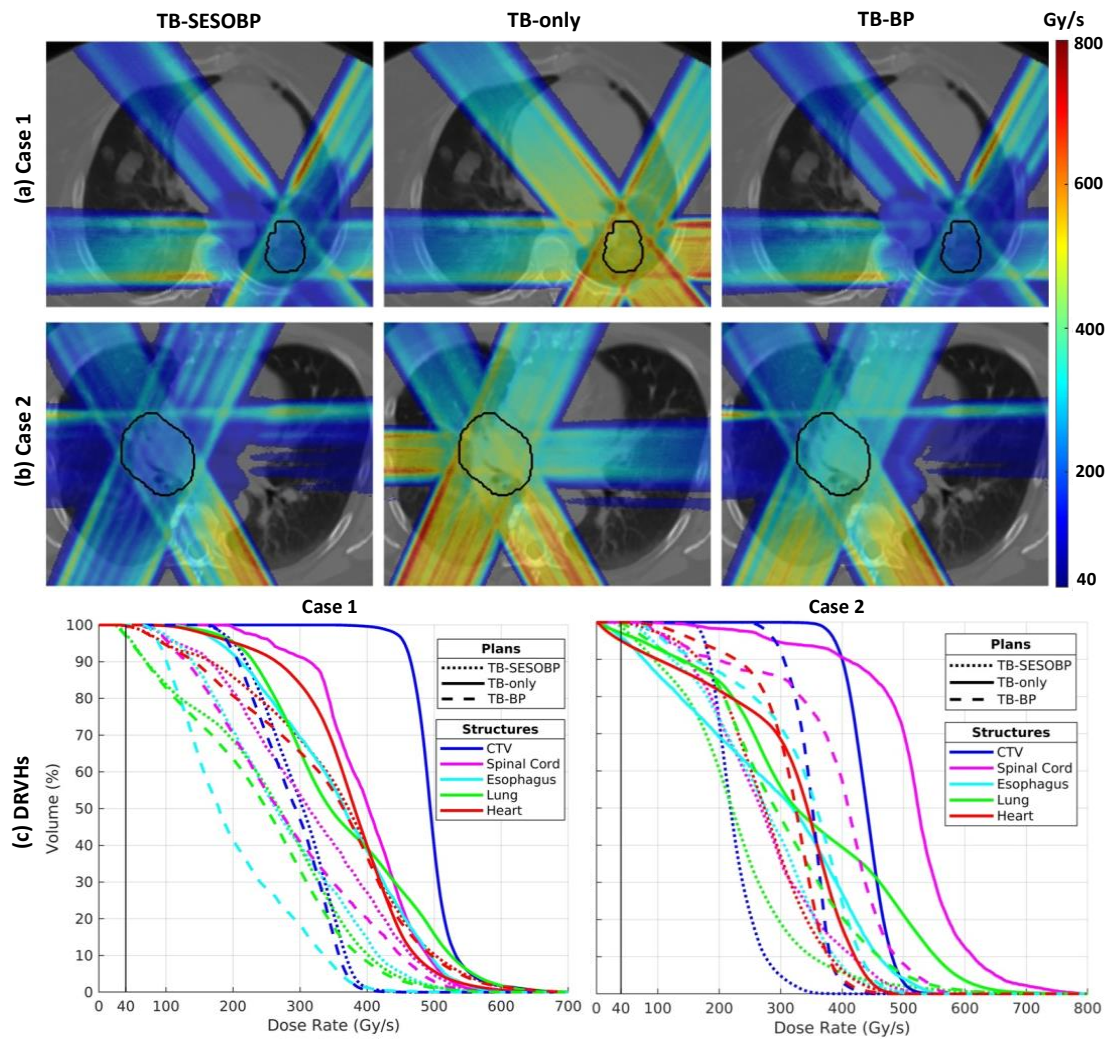

**Figure 5**. 2D dose rate distributions of the three plans for (a) Case 1 and (b) Case 2. The CTVs are contoured in black. The dose rate plot window is [40 Gy/s, 800 Gy/s]. (c) The DRVHs comparisons between the three plans for the two cases. The DRVHs of the TB-SESOBP, TB-only and TB-BP plans are marked in dotted, solid and dashed lines, respectively. A vertical line marks the 40 Gy/s threshold.

The dose rate distribution of each plan was evaluated in the volume receiving ≥10% of the prescription. According to the illustrated 2D dose rate distributions of the three plans in Figure 5 (a), the TB-only plan achieved highest dose rate in all of the dose regions, followed by the TB-BP plans. Such phenomenon was further confirmed by the extracted DRVHs of all the concerned OARs and the CTVs as are shown in Figure 5 (b). Owning to the attenuation of the RSs and RFs in generating SESOBPs or only the RSs in generating BPs, the delivered dose by the SESOBP or BP at unit weight was lower than that of the unit weighted TB. Receiving dose from more spots and lower dose per spot lead to reduced dose rate in the hybrid beam plans according to the definition of the DADR. This also implies that the highest dose rate appeared in the beam entrance of less beam interactions rather than the target of highest dose among all the structures for all of the three plans. The FLASH dose rate coverage for each structure was quantified by the percentage of the volume receiving ≥40 Gy/s, $V_{40Gy/s}$. For each TB-only plan, the $V_{40Gy/s}$ of every concerned structure reached 100%, which means that the FLASH dose rate was achieved in all the OARs and the target. As are listed in Table 4, the spinal cord and the CTV were fully covered by the FLASH dose rate in the hybrid beam plans. Besides, high and comparable FLASH dose rate coverage of the $V_{40Gy/s} \geq$ 91.4%, 97.4% and 85.5% was achieved in lung, esophagus and heart, respectively, in the hybrid beam plans. Note that no clinically significant dose was received by the heart of Case 5, therefore, the $V_{40Gy/s}$ of the heart for this case was not available. In a word, the hybrid TB-SESOBP plans dosimetrically outperformed the TB-only plans in lung and spinal cord sparing as well as the target dose homogeneity while preserving high FLASH dose coverage.

Table 4. $V_{40Gy/s}$ (%) of the TB-SESOBP, TB-only and TB-BP plans for the five cases

|  |  | Lung | Spinal cord | Esophagus | Heart | CTV |
|---|---|---|---|---|---|---|
| Case 1 | TB-SESOBP | 97.5 | 100 | 100 | 99.7 | 100 |
|  | TB-BP | 97.1 | 100 | 100 | 99.8 | 100 |
| Case 2 | TB-SESOBP | 99.0 | 100 | 100 | 100 | 100 |
|  | TB-BP | 98.7 | 100 | 100 | 100 | 100 |
| Case 3 | TB-SESOBP | 99.2 | 100 | 97.4 | 85.5 | 100 |
|  | TB-BP | 99.4 | 100 | 99.8 | 96.1 | 100 |
| Case 4 | TB-SESOBP | 92.3 | 100 | 100 | 94.2 | 100 |
|  | TB-BP | 91.4 | 100 | 100 | 94.1 | 100 |
| Case 5 | TB-SESOBP | 97.9 | 100 | 98.1 | — | 100 |
|  | TB-BP | 100 | 100 | 100 | — | 100 |

## 4. Discussion

In this study, we have proposed a hybrid inverse planning method with the combination use of the SESOBP plateau region and entrance region of the high-energy proton beams for FLASH therapy. Compared with the representative pilot study of employing TBs alone for FLASH planning[35], the cooperation of the SESOBPs and the TBs in this method could not only accomplish FLASH dose rate based on the high current of the high-energy beams, but also exploit the high LET effect and sharp dose falloff of the BPs to enhance the relative biological effect and reduce radiation exposure of the OARs at the distal edge of the targets. Additionally, the SESOBPs of FLASH dose rate in the proposed method could potentially improve tumor control while reducing lethality[21] comparing to utilizing BPs of conventional dose rate in conjunction with TBs for FLASH planning[19]. In contrast with using tailored BPs or SESOBPs of FLAHS dose rate by customized range compensators[23] or pin-shaped RFs, respectively, to obtain high LET effect, the proposed hybrid planning method is more applicable for proton ART. When replanning is required due to patient anatomical changes during the treatment course, the hybrid TB-SESOBP plan can be adapted by simply re-selecting the RFs from the pre-designed RF set if necessary and then re-conducting the optimization process based on current patient anatomy.

To our knowledge, this was the first attempt to apply the general bar RFs in proton FLASH planning using pencil beam scanning to generate SESOBPs. In order to balance the trade-off between the high computation efficiency, including memory consumption as well as computation speed in dose calculations, and the accurate simulation of the optimized RF structures, a relatively low resolution of 1 mm was used in translating the optimal weights of RF steps to the corresponding step widths. Unsurprisingly, the SOBPs of relatively high fluctuation as presented in section 3.1 were resulted from the final designed RFs. However, such dose inhomogeneity can be eliminated with the collaboration of the multiple-field intensity optimization in the proposed hybrid beam planning. Coupled with TBs, even higher dose homogeneity was accomplished in the targets by the hybrid TB-SESOBP plans comparing to the TB-only plans due to the increased degrees of freedom in the optimization process. Attributing to the sharp dose falloff of the SESOBPs, statistically significant improvements were acquired in the lungs and spinal cord sparing by the hybrid TB-SESOBP plans over the TB-only plans. In addition, the hybrid TB-SESOBP planning was more beneficial to the cases of relatively large targets, such as Case 2 and 3 in this study, than the hybrid TB-BP planning. With a more significant part of the target covered by the SOBP plateau region, the proportion of the TB component to maintain high dose homogeneity at the central target decreased, thereby better protection could be obtained for the OARs beyond the distal edge of the target. Therefore, the proposed hybrid TB-SESOBP planning method has great potential in bringing good OAR protection while maintaining high target dose homogeneity for proton FLASH therapy.

As aforementioned, the proton intensity decreases with the involvement of the RSs and RFs in beam delivery, which can lead to dose rate reduction. Thus, the TBs in the hybrid beam planning played an essential role in boosting dose rate besides maintaining high target dose homogeneity. The results indicated that high FLASH effect was achieved by the hybrid beam plans of all the target volume and at least 85% of the OAR volume receiving clinically significant dose were covered by the FLASH dose rate. The accomplishment of such high FLASH dose rate coverage was also attributed to the high prescription dose of 34 Gy in one fraction employed in this study which is recommended by the RTOG 0915 protocol[33] and corresponded to high proton intensity in beam delivery. Besides, using fewer fields in planning could be another key factor to ensure high FLASH dose coverage. Considering the trade-off between the FLASH dose coverage and the plan quality, we chose a relatively low beam current of 165 nA in FLASH mode and the lowest achievable minimum spot duration time of 0.5 ms in planning, which allowed a higher minimum MU/spot corresponding to $5.15 \times 10^8$ protons and thus more flexibility in planning for higher plan quality. It is worth noting that the dose rate in this work was quantified by DADR which only considered the instantaneous dose rates and may overestimated the dose-rate effect by ignoring dead times between spots[36]. However, the impact of the dead times between spots on the FLASH effect remains unclear. Further biological evidence is needed to help building appropriate dose rate metrics for assessing the FLASH effect.

The limitations of the current work on selecting the RFs and RSs for the hybrid TB-SESOBP planning should be noted. At each beam angle, the thickness of the RS and the height of the general bar RF were empirically determined by the 90th percentile WEDs of the distal edges and the 10th percentile WEDs of the proximal edges of the target, which may not be optimal especially when large variabilities in the WEDs of the distal and/or proximal edges exist. To fully exploit the potential of the general bar RFs in FLASH planning, this issue will be addressed in our future study by formulating an optimization problem with the thickness of the RS and the height of the bar RF as the variables. By solving this problem, the optimal parameters of the RF and RS can be found to maximize the overlapped area of the SOBP plateau region with the target and minimize the area of the SOBP plateau region beyond the target at a certain beam angle. More benefits could be gained from taking the overlapping of the SOBP plateau region with the target and the surrounding normal structures into consideration in choosing the beam angles for the hybrid TB-SESOBP planning.

The resolution issues in simulating the RFs cannot be ignored. To involve the RFs in dose calculation using MCsquare, the RF structures were added in each CT scan and shared the same voxel size with the patient anatomy. Thus, a relatively low resolution of 1 mm (typically 0.1 mm) was used in the design of the step-shaped RF bars to assure the efficiency of dose calculation, which resulted in fluctuated SOBP plateau regions, especially the plateau regions of large widths ($\geq 5.5$ cm). Likewise, the widths of ridge bars were 10-11 mm in this feasibility study of the hybrid TB-SESOBP planning, which is relatively large comparing to the conventional RF bar of width 2-5 mm[21,26], for the purpose of preserving the optimized RF structures as much as possible at such low resolution. As a consequence, the total number of the RF bars in each general RF was reduced at least by half, and the modulation ability could be degraded. Nevertheless, how the density of the RF bar can affect the modulation is still unclear and deserves further investigations. The resolution related issues could be tackled in the future study by employing an alternative Monte Carlo code of more flexibilities, such as Geant4, which allows simulating the particle transportations in the RF and the patient anatomy separately. In this way, a general RF of high resolution and densely distributed ridge bars could be used in simulating the particle transporting through the RF and generating a phase space (PS) file to record the kinetic energy, position and direction of each particle at a transverse plane downstream the RF, which could then be repeatedly utilized as the source for dose calculation in patient anatomy at a larger and clinically acceptable voxel size (e.g., 3 mm). With the highly preserved beam modulation power of the general bar RFs designed at a high resolution, we believe that the demand of the TB component for dose compensation will be reduced in the hybrid TB-SESOBP planning and the plan quality will be further improved accordingly.

## 5. Conclusions

In this study, we have demonstrated that the hybrid inverse planning with TBs and SESOBPs for proton FLASH therapy was feasible to achieve FLASH dose rate for proton therapy. Generating the SESOBPs by pre-designed general bar RFs, the proposed hybrid TB-SESOBP planning method can be easily implemented for adaptive radiotherapy. As an alternative FLASH planning approach to TB-only planning, the hybrid TB-SESOBP planning has great potential in improving OAR sparing while maintaining high target dose homogeneity.

**Conflicts of interest**

The authors have no conflicts to disclose.

**Acknowledgment**

This research is supported in part by the National Cancer Institute of the National Institutes of Health under Award Number R01CA215718 and R01EB032680.